\documentclass[prl,aps, twocolumn]{revtex4}
\usepackage{epsfig}
\usepackage{amsmath}
\input{epsf}
\usepackage{psfrag}

\usepackage[usenames,dvipsnames]{color}

\begin{document}


\vspace*{2cm}
\title{ Probing the gluon tomography  in photoproduction of di-pions}

\author{Yoshikazu Hagiwara}
 \affiliation{\normalsize\it Key Laboratory of
Particle Physics and Particle Irradiation (MOE),Institute of
Frontier and Interdisciplinary Science, Shandong University,
(QingDao), Shandong 266237, China }

\author{Cheng~Zhang}
 \affiliation{\normalsize\it Key Laboratory of
Particle Physics and Particle Irradiation (MOE),Institute of  
Frontier and Interdisciplinary Science, Shandong University,
(QingDao), Shandong 266237, China }

\author{Jian~Zhou}
 \affiliation{\normalsize\it Key Laboratory of
Particle Physics and Particle Irradiation (MOE),Institute of
Frontier and Interdisciplinary Science, Shandong University,
(QingDao), Shandong 266237, China }

\author{Ya-jin Zhou}
\affiliation{\normalsize\it Key Laboratory of Particle Physics and
Particle Irradiation (MOE),Institute of Frontier and
Interdisciplinary Science, Shandong University, (QingDao), Shandong
266237, China }

\begin{abstract}
A sizable $\cos 4\phi$ azimuthal asymmetry in exclusive di-pion production near $\rho^0$ resonance peak in ultraperipheral heavy ion collisions  recently has been  reported by STAR collaboration. We show that both elliptic gluon Wigner distribution and final state soft photon  radiation  can give rise to  this azimuthal asymmetry. The fact that the QED effect alone severely underestimates the observed asymmetry   might signal the existence of the nontrivial correlation in quantum phase distribution of gluons.
\end{abstract}

\maketitle

{\it Introduction. }
The study of nucleon/nucleus  multiple dimensional  partonic structure has been widely recognized as one of the most important frontiers of hadronic physics. Within perturbative QCD factorization framework, 3D partonic structure of nucleon  is  quantified by   transverse momentum dependent distributions(TMDs) and generalized parton distributions(GPDs). The mother distribution of these two, namely quantum phase space Wigner distribution later was introduced~\cite{Belitsky:2003nz} to accommodate more comprehensive  information on nucleon internal structure, such as parton canonical orbital angular momentum~\cite{Lorce:2011kd,Hatta:2011ku,Ji:2012sj,Hatta:2016aoc}, the quantum correlation between  the position and momentum of partons. The Fourier transform of the Wigner distribution known as the generalized TMD distributions(GTMDs)~\cite{Meissner:2009ww,Meissner:2008ay} is also often used in phenomenological studies. Pioneered by a recent work~\cite{Hatta:2016dxp}, a few proposals for extracting  the Wigner distribution in hard scattering processes in ep/A collisions for gluon sector~\cite{Hatta:2016dxp,Altinoluk:2015dpi,Zhou:2016rnt,Hagiwara:2017fye,Iancu:2017fzn,Boussarie:2018zwg,Hatta:2017cte,Mantysaari:2019csc,Mantysaari:2019hkq,Mantysaari:2020lhf,Dumitru:2021mab,Bhattacharya:2018lgm} and pp collisions for quark sector~\cite{Bhattacharya:2017bvs}   have been put forward.

One special gluon Wigner distribution  commonly referred to as the elliptic gluon Wigner distribution~\cite{Hatta:2016dxp} has recently attracted much attention~\cite{Altinoluk:2015dpi,Zhou:2016rnt,Hagiwara:2017fye,Iancu:2017fzn,Boussarie:2018zwg,Hatta:2017cte,Mantysaari:2019csc,Mantysaari:2019hkq,Mantysaari:2020lhf,Dumitru:2021mab}. It describes an  azimuthal correlation between the impact parameter and gluon transverse momentum.  In the small $x$ limit, this gluon distribution can be dynamically generated either from a semi-classical model or from small $x$ evolution. In the kinematic region where nuclear recoil transverse momentum and gluon transverse momentum are comparable, the MV model calculation predicates a rather sizable correlation~\cite{Mantysaari:2020lhf}, which we view as a very encouraging result. This distribution is proposed to be extracted via a $\cos 2\phi$ azimuthal asymmetry in diffractive di-jet production  in ep/A collisions. To avoid   the contributions from the final state soft gluon radiation  to the asymmetry~\cite{Catani:2014qha,Catani:2017tuc,Hatta:2020bgy,hatta2021azimuthal,CMS:2020ekd}, it is necessary to directly measure nuclear recoil transverse momentum instead of the total transverse momentum of di-jet, which imposes a big challenge to experimental measurement.

On the other hand, among many exciting topics~\cite{Baur:1998ay,Klein:2020fmr,Adam:2020mxg,Hatta:2019lxo,Hatta:2019ocp,Steinberg:2021lfm,Hattori:2020htm,Copinger:2020nyx,Strikman:2008zz}, ultraperipheral heavy ion collisions(UPCs) provide us unique chances to investigate partonic structure of nucleus before the advent of the EIC era. In particular, with the discovery of linear polarization of coherent photons~\cite{Li:2019yzy,Li:2019sin,Xiao:2020ddm,Adam:2019mby,daniel:2019}, a new experimental avenue is opened to  study nucleus structure via  the novel polarization dependent observables in photonuclear reactions, which already led to fruitful results~\cite{Xing:2020hwh,Zha:2020cst,Brandenburg:2021lnj,Hagiwara:2020juc}.  By coupling with elliptic gluon distribution,  linearly polarized coherent photon distribution also plays an important  role in inducing  the  $\cos 4\phi$ asymmetry in exclusive $\pi^+$ $\pi^-$ pair production in ultraperipheral heavy ion collisions recently reported by STAR collaboration~\cite{daniel:2019}, where $\phi$ is  the azimuthal angle between pion's transverse momentum and the  pion pair's total transverse momentum.

 To be more specific, the  $\cos 4\phi$ asymmetry partially results from an interference contribution. Near the $\rho^0$ resonance peak, the  di-pion  is dominantly from the decay of  $\rho^0$ which is produced in the coherent photonuclear reaction. However, there is also a non-negligible contribution to the pair invariant mass spectrum from the interference contributions between direct pions production and those from the $\rho^0$ decay~\cite{Soding:1965nh,Klusek-Gawenda:2013rtu,Bolz:2014mya,Hagler:2002nh,Hagler:2002nf,Hagiwara:2020juc}. Such an interference effect not only leads to  sizable deviation from the Breit-Wigner resonance shape~\cite{Soding:1965nh}, but also generates the  mentioned $\cos 4\phi$ azimuthal modulation.  The underlying physics of the  $\cos 4\phi$ modulation from the interference effect  is that the orbital angular momentum carried by di-pion in the direct production amplitude and in the conjugate $\rho^0$ decay amplitude  differ by four units angular momentum. In the next section, we  explain in details how  this mechanism is realized through the coupling of  elliptic gluon distribution and linearly polarized photon distribution.   Though the azimuthal asymmetry can  result from the final state soft photon radiation effect~\cite{Hatta:2020bgy} as well, our numerical results suggest that the QED effect alone is not sufficient to account for the observed asymmetry.

The paper is structured as follows. We compute the contributions to the asymmetry from both elliptic gluon distribution and final state soft photon radiation  in the next section, followed by the numerical results  presented in Sec.III. The paper is summarized in Sec.IV.

{\it $\cos 4\phi$ azimuthal asymmetry in exclusive pion pair production.}
 We start by briefly reviewing the calculation of polarization independent diffractive $\rho^0$ production (for the  recent experimental studies of the process, see Refs.~\cite{Adamczyk:2017vfu,Sirunyan:2019nog,Acharya:2020sbc}). It  is conventionally formulated in  the dipole model~\cite{Ryskin:1992ui,Brodsky:1994kf}, where the whole process is divided into  three steps: quasi-real photon splitting into quark and anti-quark pair, the color dipole scattering off nucleus, and subsequently recombining  to form  a vector meson after penetrating the nucleus target. Following this picture, one can directly write the $\rho^0$ production amplitude(see Refs.~\cite{Xing:2020hwh,Hagiwara:2020juc} for more details),
 \begin{eqnarray}
 {\cal A}(x_g,\Delta_{\!\perp}) \! = \!  \!\! \int \!\!d^2
 b_{\!\perp} e^{\!-i \Delta_{\!\perp} \! \cdot b_{\!\perp}} \!\! \int \!\! \frac{d^2  r_{\!\perp}}{4\pi}  \! N(r_{\!\perp},\! b_{\!\perp}) [\Omega^*\!K](r_{\!\perp})
  \end{eqnarray}
  where   $N(r_\perp,b_\perp)$ is the dipole-nucleus scattering amplitude, and $\Delta_\perp$ is nuclear recoil transverse momentum. And $[\Omega^*\!K]$ denotes the overlap of  photon wave function and the vector meson wave function,
 \begin{eqnarray}
 && \!\!\!\!\!\!\! [\Omega^*\!K](r_{\! \perp})\! =\!\frac{N_{\!c} e e_q}{\pi}\!\! \int_0^1 \!\!\! dz\!
 \left \{m_q^2 \Omega^*(|r_\perp|,z)K_0(|r_\perp| e_f) \right .\ \\   &&  \!\!\!\!\!  \left .\ 
\!+\!  \left [  z^2\!+\!(1\!-\!z)^2   \right ]
  \!  \frac{\partial\Omega^*(|r_\perp|,z)}{\partial |r_\perp|}
  \frac{\partial  K_0(|r_\perp| e_f)}{\partial |r_\perp|} \!\right \} \! e^{i(z\!-\!\frac{1}{2})\Delta_{\!\perp} \cdot r_{\!\perp} }  \nonumber
 \end{eqnarray}
 where $z$ stands for the  fraction of photon's light-cone momentum carried by quark. Quark and antiquark helicities and color sums  have been performed in the above formula. $K_0$ is a modified  Bessel function of the second kind. $\Omega$ is the scalar part of vector meson wave function. Note that a phase factor  $ e^{i(z-\frac{1}{2} )\Delta_\perp \cdot r_\perp}$ is included to  account for the non-forward
 correction~\cite{Bartels:2003yj,Hatta:2017cte,Hagiwara:2020mqb}. 

 One  can easily derive the dipion production amplitude  by multiplying the $\rho^0$ production amplitude with  a simplified Breit-Wigner form which describes   the transition from $\rho^0$ to $\pi^+ \ \pi^-$,
 \begin{eqnarray}
 {\cal M}_{r}= i{\cal A}(x_g,\Delta_\perp)   \frac{ f_{\rho \pi \pi} P_\perp \cdot \hat k_\perp}{Q^2\!-M_\rho^2\!+iM_\rho \Gamma_\rho}
 \end{eqnarray}
where $M_\rho$ is $\rho^0$ mass. $Q$ is the invariant mass   of the dipion system and $P_\perp$ is  defined as $P_\perp=(p_{1\perp}-p_{2\perp})/2$ with $p_{1\perp}$ and $p_{2\perp}$ being the produced pions' transverse momenta. The polarization vector of the incident photon is replaced with its transverse momentum $k_\perp$ by making use of the fact that the coherent photons are naturally linearly polarized.  $f_{\rho \pi \pi}$ is the effective coupling constant and fixed to be $f_{\rho \pi \pi}=12.24$ according to the optical theorem with the parameter $\Gamma_\rho=0.156 \ {\text {GeV}}$.  With this production amplitude, it is straightforward to  obtain  the polarization independent cross section $\frac{d \sigma}{d {\cal P.S.}}$ where $ {\cal P.S.}$ represents the phase space.

The mentioned   $\cos 4\phi$ azimuthal asymmetry can be induced by soft photon radiation from the final state charged pion particles.  The corresponding physics  from the final state photon radiation is captured by the soft factor which enters the differential  cross section formula via~\cite{Catani:2014qha,Catani:2017tuc,Hatta:2020bgy,hatta2021azimuthal}, 
\begin{eqnarray}
    \frac{d \sigma(q_\perp)}{d {\cal P.S.}}=\int d^2 q_\perp' \frac{d \sigma_0(q_\perp')}{d {\cal P.S.}}
    S(q_\perp-q_\perp')
\end{eqnarray}
where $ \sigma_0 $ is the leading order Born cross section whose full expression  can be found in Ref.~\cite{Xing:2020hwh,Hagiwara:2020juc}. The soft factor is expanded at the leading order as~\cite{hatta2021azimuthal}, 
\begin{eqnarray}
S(l_{ \perp})\!=\! \delta(l_{ \perp})\!+\! \frac{\alpha_e }{\pi^2 l_{ \perp}^2} \left \{ c_0\!+\!2 c_2 \cos 2\phi\!+\!2 c_4\cos 4\phi+... \right \} 
\label{inte}
\end{eqnarray}
where $\phi$ is the angle between $P_\perp$ and soft photon transverse momentum  $-l_{ \perp}$.   When the final state particle mass is much smaller than $P_\perp$, there exists the analytical expressions for the coefficients $c_0\approx \ln \frac{Q^2}{m_\pi^2}$, $c_2\approx \ln \frac{Q^2}{m_\pi^2}+\delta y \sinh \delta y-2\cosh^2 \frac{\delta y}{2} \ln [2(1+\cosh \delta y)]$... where $Q$ is the invariant mass of the di-pion system and $\delta y=y_1-y_2$ is the difference between two pions' rapidities. For the current case, the mass of the final state charged particles is the same order of the invariant mass. We thus  proceed numerically  and obtain $c_0\approx 3.5$, $c_2\approx 1.1$, and $c_4\approx 0.42$. The rapidity dependence of these coefficients is quite mild for RHIC kinematics and is neglected. 

Following the standard procedure, the soft factor in Eq.~\ref{inte}  can be extended  to all orders by exponentiating the azimuthal independent part to the Sudakov  factor in the transverse position space. The resummed cross section takes the form~\cite{Catani:2014qha,Catani:2017tuc,Hatta:2020bgy,hatta2021azimuthal},
  \begin{eqnarray}
  \frac{d\sigma(q_\perp)}{d {\cal P.S.} }\!&=&\!\! \!\int \!
  \frac{d^2 r_\perp}{(2\pi)^2}
  \left [1-\frac{2\alpha_e c_2 }{\pi} \cos 2\phi_r +\frac{\alpha_e c_4}{\pi} \cos 4\phi_r\right ]  \nonumber \\ &\times& \!\! e^{i r_\perp \cdot q_\perp} e^{- Sud(r_\perp)} \!\! \int d^2 q_\perp'
  e^{i r_\perp \cdot q_\perp'}  \frac{d\sigma(q_\perp')}{d {\cal P.S.} }
  \end{eqnarray}
Here $\phi_r$ is the angle between $r_\perp$ and  $P_\perp$. In the following  numerical estimation, we took into account power corrections from the soft factor following the method outlined in Ref.~\cite{hatta2021azimuthal}.
 The Sudakov factor at one loop is given by~\cite{hatta2021azimuthal},
  \begin{eqnarray}
  Sud(r_\perp)=
\frac{\alpha_e}{\pi}c_0  {\rm ln}\frac{P_\perp^2}{\mu_r^2}   
  \end{eqnarray}
  with $\mu_r=2 e^{-\gamma_E}/|r_\perp|$.
The Sudakov factor plays a critical role in yielding  perturbative high $q_\perp$ tail of lepton pair produced  in UPCs~\cite{Klein:2018fmp,Xiao:2020ddm}.

 We now turn to derive the direct di-pion production amplitude. In the collinear factorization, the transition from quark-antiquark  pair to dipion pair is described by the nonperturbative and universal dipion distribution amplitude(DA)~\cite{Diehl:1998dk,Polyakov:1998ze,Diehl:2000uv}.
 For the current case, it is necessary to go beyond the collinear factorization and take into account quark and antiquark finite transverse momenta (or transverse position) effects in order to preserve the information of the orbital  angular momentum, which is essential for yielding the non-trivial azimuthal modulation.  To this end, we assume that the transverse position of $\pi^+$ is the same as the leading $u$ or $\bar d$ quark during this non-perturbative transition, though the longitudinal momentum  is re-distributed during the transition. The similar  requirement applies to $\pi^-$ as well. We notice that our assumption for the transverse spatial distribution differs from  the analysis made in Ref.~\cite{Pire:2002ut}. This difference might be attributed to the fact that quark and anti-quark transverse momenta have been integrated out in the work~\cite{Pire:2002ut}, while transverse momenta of quark and the corresponding pion are implicitly assumed to be the same in our treatment.    By making this crude approximation, the amplitude reads,
\begin{eqnarray}
  {\cal M}_{ d}&=&i \!\int \! d^2 b_\perp e^{i\Delta_\perp \cdot b_\perp}
  \!\int \! \frac{d^2 r_\perp}{4\pi}\int d z
 \nonumber \\ && \!\!\!\! \times 
 \Psi^{\gamma\rightarrow q\bar q}(r_\perp,z,\epsilon_{\perp \gamma}) N(r_\perp,b_\perp) \Phi(z,\zeta) e^{ir_\perp \cdot P_\perp}
 \label{4}
  \end{eqnarray}
where $\Phi$ is a universal nonperturbative function describing the transition from a quark anti-quark pair to a di-pion system. $\zeta$  is the longitudinal momentum fraction  carried by $\pi^+$. We stress that any non-perturbative model for the quark-pair to pion-pair transition   would generate a $\cos 4\phi$ azimuthal asymmetry, as long as it conserves angular momentum. 
  
We proceed to explicitly work out the angular structure residing in ${\cal M}_d$ by first writing  the polarization dependent photon wave function, 
\begin{eqnarray}
&& \!\!\!\!\!\!\!\!\!\!\! \Psi^{\gamma \rightarrow q\bar q}
(r_\perp,z,\epsilon_\perp^\gamma) =\frac{e e_q}{2\pi} \delta_{ij}
\left \{   \delta_{\sigma \sigma'}m_q(\epsilon_\perp^{\gamma,1}+i\sigma \epsilon_\perp^{\gamma,2})\right .\
\nonumber \\&& \left .\ \!\!\!\!\!\! \!\!\!\!\! -\delta_{\sigma,- \sigma'}
\!\left [ (1\!-\! 2z)i
  \epsilon_{\!\perp}^\gamma \!\! \cdot \hat r_{\!\perp} \! +\sigma \epsilon_{\!\perp}^\gamma \! \times \hat r_{\!\perp}\right ]\!
  \frac{\partial}{\partial |r_{\!\perp}|}\!
  \right \}\!
 K_0(|r_{\!\perp}| e_f)
\end{eqnarray}
where  $\sigma$ and $\sigma'$ are the quark and antiquark helicities, i, j are the color indices. And $\epsilon_{\perp}^\gamma=\hat k_\perp \equiv k_\perp/|k_\perp|$ as explained above.  $m_q$ and $e_q$ denote  quark mass and quark's electric charge number with flavor $q$ respectively.
 And $e_f$ is defined as $e_f^2=Q^2z(1-z)+m_q^2$ for the virtual photon case. For the quasi-real photon case, $e_f=140$MeV is determined by fitting to HERA data.

 We proceed to discuss  the symmetric properties of $\Phi(z,\zeta)$ in order to have some  guidance  for  parameterizing this  non-perturbative function. The invariance of the scattering amplitude under the charge parity transformation requires 
 \begin{eqnarray}
 \Phi(z,\zeta) = -\Phi(1-z,1-\zeta) \ . 
  \end{eqnarray}
On the other hand, the final state produced  di-pion is a C-odd state since the orbital angular momentum quantum number carried by di-pion is either 1 or 3 in the current case.  This implies that 
\begin{eqnarray}
  \Phi(z,\zeta)e^{ir_\perp \cdot P_\perp} = -\Phi(z,1-\zeta)e^{-ir_\perp \cdot P_\perp}  \  .
   \end{eqnarray}
Using these relations, one obtains,
\begin{eqnarray}
  \Phi(z,\zeta)e^{ir_\perp \cdot P_\perp}& =& \!\! \frac{ \Phi(z,\zeta)+\Phi(z,1-\zeta)  }{2} e^{ir_\perp \cdot P_\perp}\nonumber \\&-&\!\!  \Phi(z,1-\zeta)\frac{ e^{-ir_\perp \cdot P_\perp}+e^{ir_\perp \cdot P_\perp}}{2}
   \end{eqnarray}
   where the contribution in the second line  decouples after convoluting with the photon wave function and the dipole amplitude in Eq.~\ref{4}. More discussions  on the symmetric properties for  the integrated di-pion DA can be found in  Ref.~\cite{Diehl:2003ny}.
We are then  motivated  to parametrize the    $z$ integration in Eq.~\ref{4} as the following,
 \begin{eqnarray}
  \int \!\! dz (1-2z) \Phi(z,\zeta) \!\!&=& \!\!
  \int  \!\! dz (1-2z)  \frac{ \Phi(z,\zeta)+\Phi(z,1-\zeta)  }{2}\nonumber \\&=& \!\!\delta_{\sigma,- \sigma'} \delta^{ij} \zeta(1-\zeta) {\cal C}
  \end{eqnarray}
 where $\zeta(1-\zeta)$ is a commonly used parametrization that respects particle-antiparticle symmetry. The constant ${\cal C}$ is determined by fitting  to the polarization averaged  cross section which receives the sizable contribution from the interference term.  We close the discussion about dipion distribution amplitude with a final remark.  $\Phi(z,\zeta) $ does not correspond  to the normal leading twist collinear dipion distribution amplitude. Actually, the polarization independent leading twist dipion DA decouples from the process under consideration once averaging  over the polarizations of the incident photon.  Instead, the high twist collinear DAs that contribute to the observable are related to the functions  $  \frac{\partial }{\partial r_\perp^i} \Phi(z,\zeta)e^{ir_\perp \cdot P_\perp}|_{r_\perp \rightarrow 0}$ for the p wave final state and $ \frac{\partial}{\partial r_\perp^i} \frac{\partial}{\partial r_\perp^j}\frac{\partial}{\partial r_\perp^k} \Phi(z,\zeta)e^{ir_\perp \cdot P_\perp}|_{r_\perp \rightarrow 0}$ for the f wave final state.  Here we refrain ourself from giving the explicit matrix element definition for these  di-pon DAs as we do not attempt to extensively discuss the properties of high twist dipion DAs in this work.

We now study the angular correlation in the dipole amplitude $N(b_\perp, r_\perp)$ which reads in the MV model~\cite{McLerran:1993ka,McLerran:1993ni},
\begin{eqnarray}
N (b_\perp, r_\perp)\!
=\!1- \!e^{\! -\alpha_s  C_F \frac{1}{8\pi} \! \int \! d^2x_{\! \perp} \mu_A(x_\perp)
  \left [ {\rm ln} \frac{(b_\perp+\frac{r_\perp}{2}-x_\perp)^2}{(b_\perp-\frac{r_\perp}{2}-x_\perp)^2}
 \right ]^2 }
 \label{7}
\end{eqnarray}
where $\mu_A(x_\perp)$ is the color source distribution. If $\mu_A(x_\perp)$ is not uniformly distributed in the transverse plane of nucleus, a nontrivial angular correlation between $b_\perp$ and $r_\perp$ can arise.  By Taylor expanding the above expression and ignoring high order harmonics, one obtains,
\begin{eqnarray}
&& \!\!\!\!\!\!\!\!\!\!\!
N(b_\perp, r_\perp) \approx 1- \exp \left [ -Q_s^2 \left (b_\perp^2 \right ) r_\perp^2/4\right ] 
\label{functionE} \\ && 
+E  \left (b_\perp^2, r_\perp^2 \right ) 2 \cos(2\phi_b-2\phi_r)\frac{Q_s^2 \left (b_\perp^2 \right ) r_\perp^2}{4} e^{ -\frac{Q_s^2 \left (b_\perp^2 \right ) r_\perp^2}{4}} 
\nonumber
\end{eqnarray}
The coefficient $ E \left (b_\perp^2, r_\perp^2 \right )$ has been explicitly computed in the MV model for different kinematic regions~\cite{Zhou:2016rnt,Iancu:2017fzn,Mantysaari:2019hkq,Mantysaari:2020lhf}. In this work, it is treated as a free parameter since  the kinematics of interest is in the non-perturbative region. 

At this step, one can carry out the integration over the azimuthal angles of $r_\perp$ and $b_\perp$. The direct production amplitude is re-written as,
\begin{eqnarray}
 {\cal M}_{ d}\!\! &=& \!\!  i \zeta(1-\zeta) \cos (2\phi_\Delta+\phi_k-3\phi_P){\cal E}(x_g, \Delta_\perp)  \nonumber \\ \!\!  &+& \!\!  i \zeta(1-\zeta) \cos (\phi_k-\phi_P){\cal A}_{d}(x_g, \Delta_\perp)+...
\end{eqnarray}
 where we only keep the angular structures which contribute to the azimuthal independent interference cross section and the  $\cos 4\phi$ azimuthal modulation. $\phi_k$, $\phi_P$ and $\phi_\Delta$ represent the azimuthal angles of different transverse momenta respectively. ${\cal E}(x_g, \Delta_\perp)$ and ${\cal A}_{d}(x_g, \Delta_\perp)$ are given by,
  \begin{eqnarray}
    \!\!\!\!\!\!\! \! 
    {\cal E}(x_g, \Delta_\perp)\!\!&=&\!\!- \frac{ee_q}{2} N_c {\cal C} Ee^{i \delta \phi} \!\int \! \frac{d b_\perp^2 dr_\perp^2}{4} \frac{Q_s^2 r_\perp^2}{4} e^{ -\frac{Q_s^2  r_\perp^2}{4}} \label{E} \nonumber \\ 
  &\times&  \!\!\!
    J_2(|\Delta_\perp||b_\perp|) J_3(|r_\perp||P_\perp| )  \frac{\partial K_0(|r_\perp|e_f)}{\partial |r_\perp|} \\ 
    \!\!\!\!\!\!\! \! 
    {\cal A}_d(x_g, \Delta_\perp)\!\!&=&\!\!-ee_qN_c {\cal C}  \!\int \! \frac{d b_\perp^2 dr_\perp^2}{4} \left ( 1-e^{ -\frac{Q_s^2  r_\perp^2}{4}}\right )\nonumber \\
    &\times&  \!\!\!
    J_0(|\Delta_\perp||b_\perp|) J_1(|r_\perp||P_\perp| )  \frac{\partial K_0(|r_\perp|e_f)}{\partial |r_\perp|} 
\end{eqnarray}
where  ${\cal A}_d$ contributes to the polarization independent interference cross section, which will be  compared with STAR data to fix the constant $\cal C$. As there is no experimental evidence suggesting that the azimuthal dependent amplitude must be real, we insert a phase on the right hand side of Eq.~\ref{E}. To have a maximal asymmetry, we assume $\delta \phi=\pi/2$ in the following numerical calculation.

Since the two incoming nuclei take turns to act as the coherent photon source and  the target, one should sum over these two possibilities on the amplitude level.    The observed diffractive pattern at low transverse momentum crucially depends on such double slit  interference effect~\cite{Klein:1999gv,Abelev:2008ew,Zha:2018jin,Xing:2020hwh,Zha:2020cst}.
 We noticed that the similar double slit interference effect has been discovered in spin space~\cite{Chen:2020adz,Chen:2021gdk}.  Following the method outlined in Refs~\cite{Xing:2020hwh}, we also include the double slit interference effect in the calculation of the  interference cross section from direct production and $\rho$ decay production. The azimuthal dependent part of which is given by,
 \begin{eqnarray}
  && \!\!\!\!\!\!\!\!\!\!\!
   \frac{d \sigma_{I}}{d{\cal P.S.} } \!  = \! \frac{\zeta(1\!-\zeta)  M_\rho \Gamma_\rho |P_\perp| f_{\rho\pi\pi}}{2(2\pi)^7 ((Q^2\!-\!M_\rho^2)^2\!+\!M_\rho^2 \Gamma_\rho^2)} 
    \nonumber \\ &&\!\!\!\!\!\!\times
    \int d^2 \Delta_\perp d^2k_\perp d^2 k_\perp'
    \delta^2(k_\perp+\Delta_\perp-q_\perp) \nonumber \\ && \times 
    \cos(3\phi_{P}-\phi_{k}-2\phi_\Delta)\cos(\phi_{P}-\phi_{k'})    \left \{  \right .\
     \nonumber \\ &&\!\!\!
     e^{i \tilde b_{\!\perp} \! \cdot (k_{\!\perp}'\!\!-k_{\!\perp})}\!   
    {\cal A}^*(x_2,\Delta_{\!\perp}') {\cal E}(x_2, \Delta_{\!\perp})
    {\cal F}(x_1,k_{\!\perp}){\cal F}(x_1,k_{\!\perp}')    
      \nonumber \\ & + &  \!\!\! e^{i \tilde b_{\!\perp} \! \cdot (\Delta_{\!\perp}'\!\!-\Delta_{\!\perp})}\!
      {\cal A}^*(x_1, \Delta_{\!\perp}') {\cal E}(x_1,\Delta_{\!\perp})
    {\cal F}(x_2,k_{\!\perp}){\cal F}(x_2,k_{\!\perp}')
        \nonumber \\  & + & \!\! \!
 e^{i \tilde b_{\!\perp} \! \cdot (\Delta_{\!\perp}'\!\!-k_{\!\perp})}\!
     {\cal A}^*(x_2,\Delta_{\!\perp}') {\cal E}(x_1, \Delta_{\!\perp}){\cal F}(x_1,k_{\!\perp}){\cal F}(x_2,k_{\!\perp}')
        \nonumber \\  & + & \!\!\!\!  \!\!
     \left .\ e^{i \tilde b_{\!\perp} \! \cdot (k_{\!\perp}'\!\!-\Delta_{\!\perp}')} \!
     {\cal A}^*(x_1, \Delta_{\!\perp}')  {\cal E}(x_2,\Delta_{\!\perp}){\cal F}(x_2,k_{\!\perp})  {\cal F}(x_1,k_{\!\perp}')       \right \}\!
      \nonumber \\
      && \!\!\! \!\! +c.c.
\end{eqnarray}
  where  $d{\cal P.S.}=d^2 p_{1\perp} d^2 p_{2\perp} dy_1 dy_2 d^2 \tilde b_{\perp}$.  $y_1$ and $y_2$ are the final state pions'  rapidities. $k_\perp$, $\Delta_\perp$, $k_\perp'$ and $\Delta_\perp'$ are the incoming photon's transverse momenta and nucleus recoil transverse momenta in the amplitude and the conjugate amplitude respectively.  $\tilde b_\perp$ is the transverse distance between the center of the two colliding nuclei.  ${\cal F}(x,k_\perp)$ describes the probability amplitude for finding a photon carries certain momentum. The squared  ${\cal F}(x,k_\perp)$  is just the standard   photon TMD distribution $f(x,k_\perp)$. Using  the  equivalent photon  approximation, one has,
\begin{eqnarray}
  {\cal F}(x,k_\perp)=\frac{Z \sqrt{\alpha_e}}{\pi} |k_\perp|
 \frac{F(k_\perp^2+x^2M_p^2)}{(k_\perp^2+x^2M_p^2)}
\end{eqnarray}
where $x$ is constrained according to $x= \sqrt{\frac{P_\perp^2+m_\pi^2}{S}}(e^{y_1}+e^{y_2})$. $M_p$ is proton mass.  $Z$ is the nuclear charge number.  $F$ is the nuclear charge form factor parametrized with the standard Woods-Saxon distribution. Note that the incoming photon carry the different transverse momenta in the amplitude and the conjugate amplitude since we fixed $\tilde b_\perp$~\cite{Vidovic:1992ik,Klein:2018fmp,Zha:2018tlq,Klein:2020jom,Klusek-Gawenda:2020eja,Wu:2021ril}.

\begin{figure}[htpb]
\includegraphics[angle=0,scale=0.52]{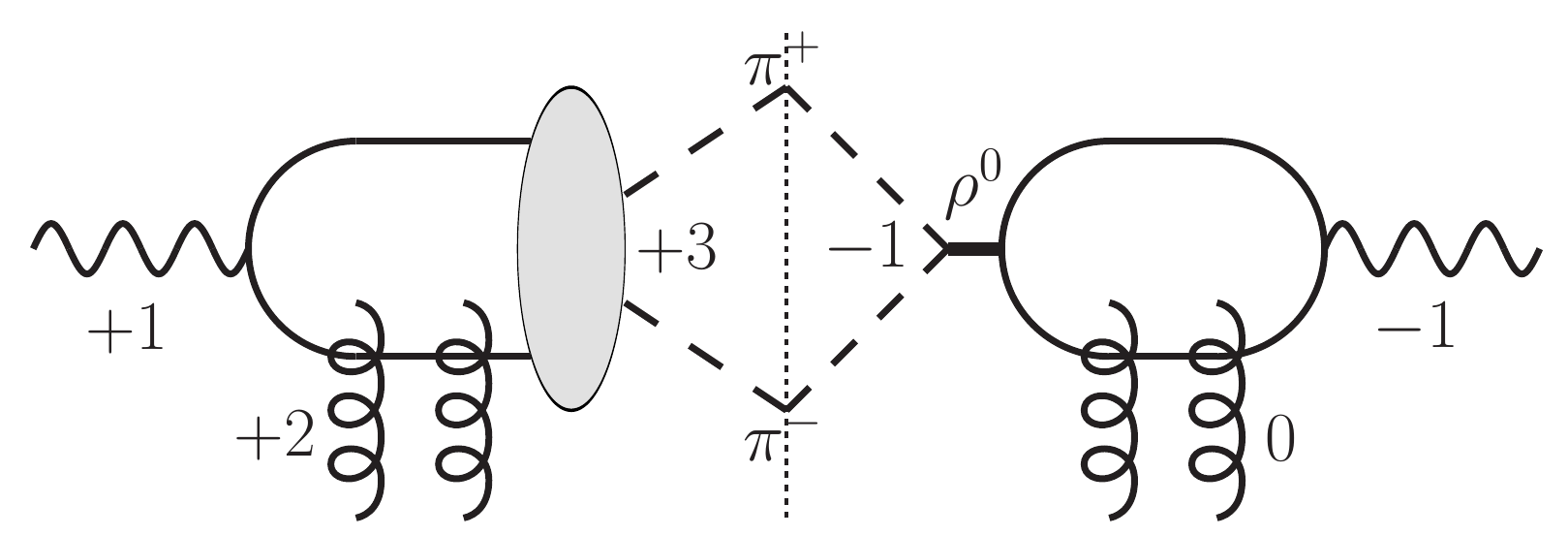}
\caption{ $\cos 4\phi$ azimuthal asymmetry results from the interference
between the p wave and the f wave of pion pairs that are from the decay of $\rho^0$ meson in conjugate amplitude, and are from direct production in the amplitude. The color neutral  exchange in the amplitude  described by the elliptic gluon distribution effectively  carries two unit orbital angular momentum. The incident photon is linearly polarized.  } \label{fig1}
\end{figure}

The emergence  of this nontrivial azimuthal correlations can be intuitively
understood as follows: the linear polarization photon state is the superposition of different helicity states. As illustrated in Fig.~\ref{fig1}, the  incoming photon  carries  1 unit spin angular momentum  in the amplitude while it carries -1 unit spin angular momentum in the conjugate amplitude. The orbital angular momentum transferred to quark pair via the two gluon exchange is 2 unit angular momentum as far as the elliptic gluon distribution is concerned.  As a result, the orbital angular momentum carried by the produced dipion system is 3 unit in the amplitude and -1 in the conjugate amplitude correspondingly. Such interference amplitude leads to a $\cos 4\phi$ angular modulation. If one averages over coherent photon polarization, this mechanism gives rise to   $\cos 2\phi$ asymmetry as well, though the dominant contribution to it comes from different source~\cite{Xing:2020hwh}. 

{ \it Numerical estimations.}
We now introduce  models/parametrizations  that are necessary for numerical calculations.
 First of all, the dipole-nucleus scattering amplitude (the azimuthal independent part)  is parametrized in terms of dipole-nucleon scattering amplitude ${\cal N}(r_\perp)$~\cite{Kowalski:2003hm,Kowalski:2006hc,Rezaeian:2012ji,Kowalski:2008sa,Kowalski:2007rw},
 \begin{eqnarray}
  N(b_\perp, r_\perp) \approx 1-\left [1-
2\pi B_p  T_A(b_\perp) {\cal N}(r_\perp)\right ]^A
  \end{eqnarray}
where   we adopt the GBW model for ${\cal N}(r_\perp)$. We also made the numerical estimates with a more sophisticated treatment for  ${\cal N}(r_\perp)$~\cite{Bartels:2002cj,Rezaeian:2012ji,Kowalski:2008sa,Kowalski:2007rw}, which leads to the similar results. The nuclear thickness function $T_A(b_\perp)$ is determined with  the Woods-Saxon distribution in our numerical calculation, and $B_p=4{\text GeV}^{-1}$. For the scalar part of  vector meson function, we use ``Gauss-LC" wave function also taken from Ref.~\cite{Kowalski:2003hm,Kowalski:2006hc}:
$\Omega^*(|r_\perp|,z)= \beta z(1-z) \exp \left[-\frac{r_\perp^2}{2R_\perp^2}\right ] $
 with $\beta=4.47$, $R_\perp^2=21.9 \text{GeV}^{-2}$.  The nuclear thickness function is estimated  with  the Woods-Saxon distribution, $F(\vec k^2)= \int d^3 r e^{i\vec k\cdot \vec r} \frac{C^0}{1+\exp{\left [(r-R_{WS})/d\right ]}}$ where  $R_{WS}$ (Au: 6.38fm) is the radius  and d (Au.:0.535fm) is the skin depth. $C^0$ is the normalization factor.

 UPCs events measured at RHIC are triggered by detecting accompanied forward neutron emissions.  
 The impact parameter dependence of the probability for emitting any number of  neutrons from an excited nucleus (referred to as the ``$X_n$" event) is described by the function, $  P(\tilde b_\perp)= 1-  \exp \left [-P_{1n}(\tilde b_\perp) \right ]$
  with $P_{1n}(\tilde b_\perp)= 5.45*10^{-5} \frac{Z^3(A-Z) }{A^{2/3} \tilde b_\perp^2 } \ \text{fm}^2 $.
Therefore, the  ``tagged" UPC cross section is defined as,
\begin{eqnarray}
2 \pi \int_{2R_A}^{\infty} \tilde b_\perp d\tilde b_\perp P^2(\tilde b_\perp) d \sigma(\tilde b_\perp, \ ...)
\end{eqnarray}
With all these ingredients, we are ready to perform numerical study of the $\cos 4\phi$ azimuthal asymmetry for RHIC kinematics.

\begin{figure}[htpb]
  \includegraphics[angle=0,scale=0.5]{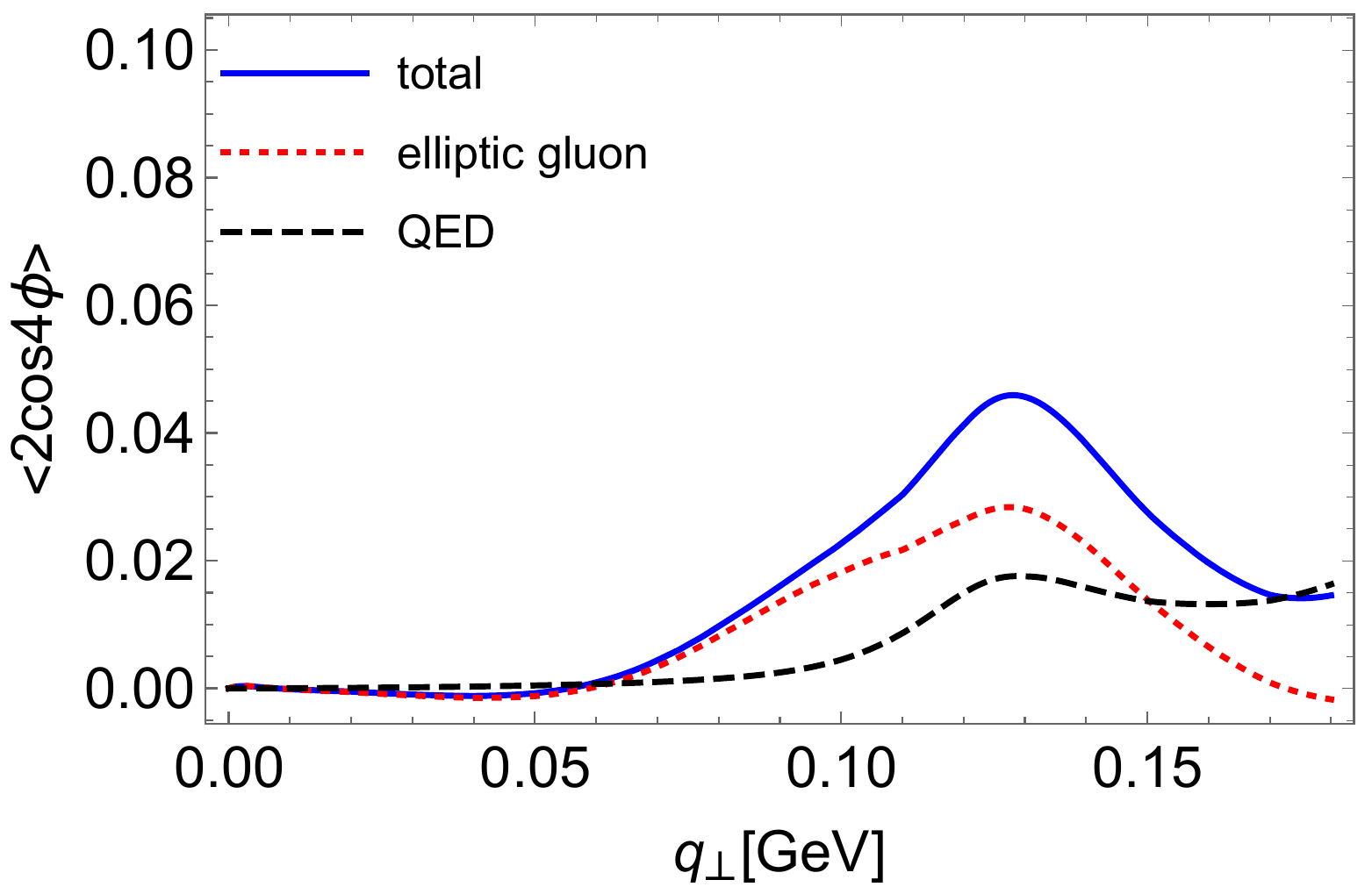}
  \caption{The asymmetry is plotted as the function of $q_\perp$ for RHIC energy $\sqrt {S}=200 $GeV.  The rapidities $y_1,y_2$ of produced pions are integrated over the region $[-1,1]$ and $Q$ is integrated over the region  $[0.6 {\text GeV}, 1{\text GeV}]$.  The contributions from the final state soft photon radiation and elliptic gluon distribution to the asymmetry are shown separately.   } \label{fig2}
  \end{figure}

We first compute  the azimuthal averaged cross section and compare it with STAR data to  fix the coefficient $ {\cal C}\approx -10$ which determines the relative magnitude  between the direct pion pair production and that via $\rho^0$ decay.  We then are able to compute the $\cos4\phi$ asymmetry from the elliptic gluon distribution.  The QED  and the elliptic gluon distribution contributions to the asymmetry are separately presented in Fig.~\ref{fig2}. If we only take into account  the final state soft photon radiation effect, the theory calculation severely   underestimates the experimental data. To match  the STAR data~\cite{daniel:2019}, a rather large value of the coefficient $E=0.4$ in the Eq.~\ref{functionE} which is roughly one order of magnitude larger than the perturbative estimate for $E$~\cite{Zhou:2016rnt,Mantysaari:2020lhf},  has been used in our numerical calculation.  Since we are dealing with the deep non-perturbative region,  it is hard to tell whether such large value for $E$ is reasonable or not. Moreover, there is a lot of uncertainties associated with the  transition from quark pair to di-pion.  Other non-perturbative model for describing this transition might lead to a much larger asymmetry with the same value of $E$.  Nevertheless, as demonstrated in Fig.~\ref{fig2}, it is clear that  the elliptic gluon  distribution is a necessary element to account for the observed asymmetry (around 10\% ). 

\begin{figure}[htpb]
    \includegraphics[angle=0,scale=0.522]{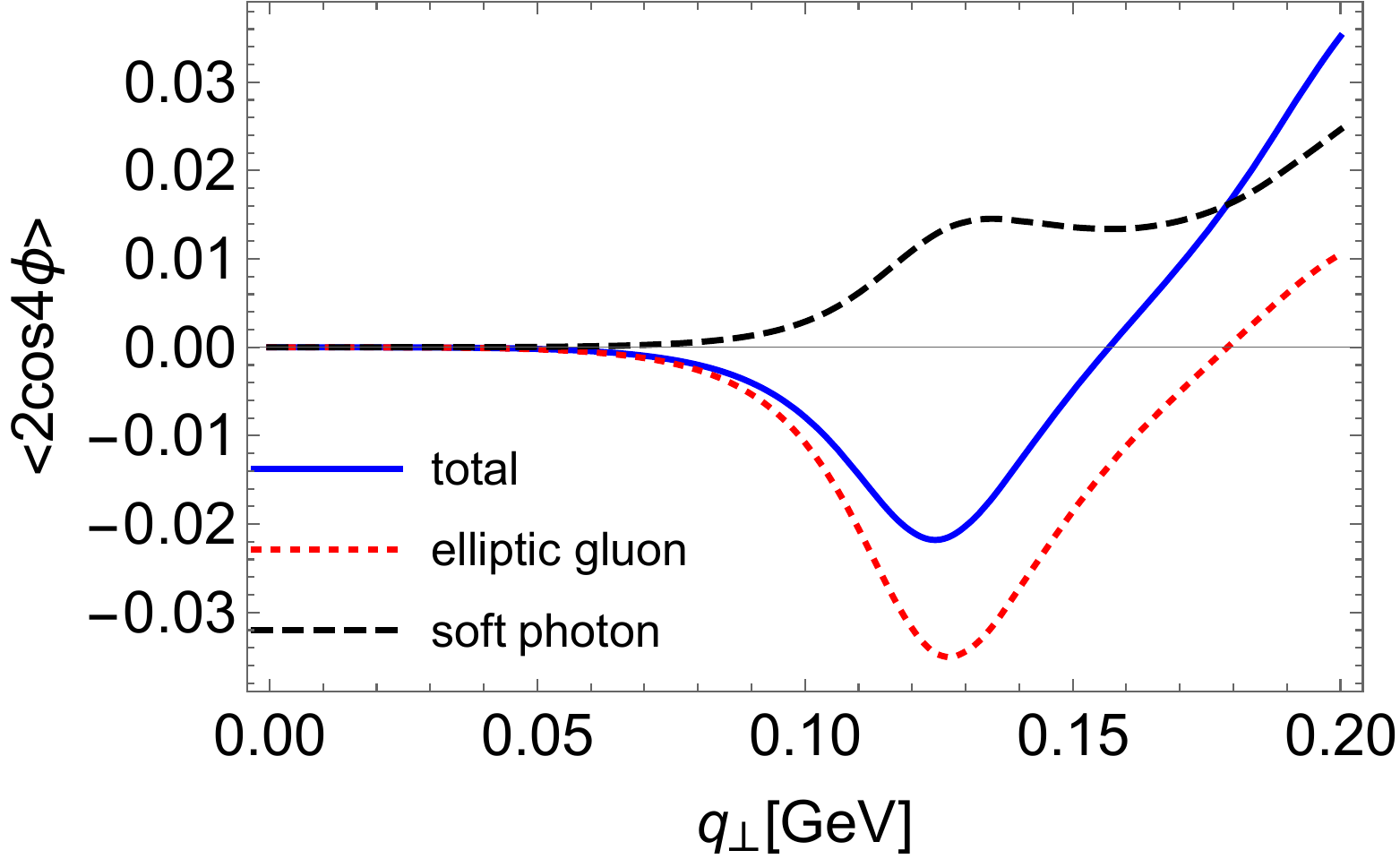}
    \caption{The asymmetry in photon production of di-pion in eA collisions at EIC is plotted as the function of $q_\perp$ for the center of mass energy $\sqrt {S}=100 $GeV.  The rapidities $y_1,y_2$ of produced pions are integrated over the region $[2,3]$ and the invariant mass of di-pion $Q$ is integrated over the region  $[0.6 {\text GeV}, 1{\text GeV}]$. Transverse momentum carried by the quasi-real photon emitted from electron beam  is required to be smaller than 0.1GeV.  } \label{fig3}
    \end{figure}

We also compute the $\cos 4\phi$ azimuthal asymmetry in the process $\gamma + A \rightarrow A'+\pi^+ +\pi^-$ for EIC kinematics with the same set parameters. It is shown in Fig.~\ref{fig3} that the contribution from the elliptic gluon distribution to the asymmetry flips the sign as the result of the absence of the double slit interference effect in eA collisions. It would be very interesting to test this predication at the future EIC. In view of the recent findings~\cite{Hatta:2020bgy,hatta2021azimuthal}, this might be the  only clean  observable to probe the gluon Wigner function at EIC, because it is free from the  contamination due to the final state soft gluon  radiation effect.

{\it Conclusion.}
We studied $\cos 4\phi$ azimuthal asymmetry in exclusive di-pion production near $\rho^0$ resonance peak in UPCs.  Both the final state soft photon radiation effect and the elliptic  gluon distribution  can give rise to such a asymmetry.   It is shown  that the QED effect alone, which can be cleanly computed, is not adequate to describe the  STAR data. On the other hand, with some model dependent input, a better agreement with the preliminary STAR data is reached   after including the elliptic gluon distribution contribution, though the theory calculation  still underestimates the measured asymmetry.  This thus leads us to conclude  that  the observed  $\cos 4\phi$ asymmetry might signal the  very existence of the non-trivial quantum correlation encoded in elliptic gluon distribution. 

{\it Acknowledgments. }
 We thank Raju Venugopalan, Zhang-bu Xu,  James Daniel Brandenburg, Feng Yuan and Chi Yang for valuable  discussions.  J. Zhou has been supported by the National Natural Science Foundation of China under Grant No.\ 11675093. Y. Zhou has been supported by the  Natural Science Foundation of Shandong Province under  Grant No.\ ZR2020MA098. C. Zhang is supported by the China Postdoctoral Science Foundation under Grant No.~2019M662317.

\bibliography{ref.bib}

\begin{thebibliography}{79}
\expandafter\ifx\csname natexlab\endcsname\relax\def\natexlab#1{#1}\fi
\expandafter\ifx\csname bibnamefont\endcsname\relax
  \def\bibnamefont#1{#1}\fi
\expandafter\ifx\csname bibfnamefont\endcsname\relax
  \def\bibfnamefont#1{#1}\fi
\expandafter\ifx\csname citenamefont\endcsname\relax
  \def\citenamefont#1{#1}\fi
\expandafter\ifx\csname url\endcsname\relax
  \def\url#1{\texttt{#1}}\fi
\expandafter\ifx\csname urlprefix\endcsname\relax\def\urlprefix{URL }\fi
\providecommand{\bibinfo}[2]{#2}
\providecommand{\eprint}[2][]{\url{#2}}

\bibitem[{\citenamefont{Belitsky et~al.}(2004)\citenamefont{Belitsky, Ji, and
  Yuan}}]{Belitsky:2003nz}
\bibinfo{author}{\bibfnamefont{A.~V.} \bibnamefont{Belitsky}},
  \bibinfo{author}{\bibfnamefont{X.-d.} \bibnamefont{Ji}}, \bibnamefont{and}
  \bibinfo{author}{\bibfnamefont{F.}~\bibnamefont{Yuan}},
  \bibinfo{journal}{Phys. Rev. D} \textbf{\bibinfo{volume}{69}},
  \bibinfo{pages}{074014} (\bibinfo{year}{2004}), \eprint{hep-ph/0307383}.

\bibitem[{\citenamefont{Lorce and Pasquini}(2011)}]{Lorce:2011kd}
\bibinfo{author}{\bibfnamefont{C.}~\bibnamefont{Lorce}} \bibnamefont{and}
  \bibinfo{author}{\bibfnamefont{B.}~\bibnamefont{Pasquini}},
  \bibinfo{journal}{Phys. Rev. D} \textbf{\bibinfo{volume}{84}},
  \bibinfo{pages}{014015} (\bibinfo{year}{2011}), \eprint{1106.0139}.

\bibitem[{\citenamefont{Hatta}(2012)}]{Hatta:2011ku}
\bibinfo{author}{\bibfnamefont{Y.}~\bibnamefont{Hatta}},
  \bibinfo{journal}{Phys. Lett. B} \textbf{\bibinfo{volume}{708}},
  \bibinfo{pages}{186} (\bibinfo{year}{2012}), \eprint{1111.3547}.

\bibitem[{\citenamefont{Ji et~al.}(2012)\citenamefont{Ji, Xiong, and
  Yuan}}]{Ji:2012sj}
\bibinfo{author}{\bibfnamefont{X.}~\bibnamefont{Ji}},
  \bibinfo{author}{\bibfnamefont{X.}~\bibnamefont{Xiong}}, \bibnamefont{and}
  \bibinfo{author}{\bibfnamefont{F.}~\bibnamefont{Yuan}},
  \bibinfo{journal}{Phys. Rev. Lett.} \textbf{\bibinfo{volume}{109}},
  \bibinfo{pages}{152005} (\bibinfo{year}{2012}), \eprint{1202.2843}.

\bibitem[{\citenamefont{Hatta et~al.}(2017{\natexlab{a}})\citenamefont{Hatta,
  Nakagawa, Yuan, Zhao, and Xiao}}]{Hatta:2016aoc}
\bibinfo{author}{\bibfnamefont{Y.}~\bibnamefont{Hatta}},
  \bibinfo{author}{\bibfnamefont{Y.}~\bibnamefont{Nakagawa}},
  \bibinfo{author}{\bibfnamefont{F.}~\bibnamefont{Yuan}},
  \bibinfo{author}{\bibfnamefont{Y.}~\bibnamefont{Zhao}}, \bibnamefont{and}
  \bibinfo{author}{\bibfnamefont{B.}~\bibnamefont{Xiao}},
  \bibinfo{journal}{Phys. Rev. D} \textbf{\bibinfo{volume}{95}},
  \bibinfo{pages}{114032} (\bibinfo{year}{2017}{\natexlab{a}}),
  \eprint{1612.02445}.

\bibitem[{\citenamefont{Meissner et~al.}(2009)\citenamefont{Meissner, Metz, and
  Schlegel}}]{Meissner:2009ww}
\bibinfo{author}{\bibfnamefont{S.}~\bibnamefont{Meissner}},
  \bibinfo{author}{\bibfnamefont{A.}~\bibnamefont{Metz}}, \bibnamefont{and}
  \bibinfo{author}{\bibfnamefont{M.}~\bibnamefont{Schlegel}},
  \bibinfo{journal}{JHEP} \textbf{\bibinfo{volume}{08}}, \bibinfo{pages}{056}
  (\bibinfo{year}{2009}), \eprint{0906.5323}.

\bibitem[{\citenamefont{Meissner et~al.}(2008)\citenamefont{Meissner, Metz,
  Schlegel, and Goeke}}]{Meissner:2008ay}
\bibinfo{author}{\bibfnamefont{S.}~\bibnamefont{Meissner}},
  \bibinfo{author}{\bibfnamefont{A.}~\bibnamefont{Metz}},
  \bibinfo{author}{\bibfnamefont{M.}~\bibnamefont{Schlegel}}, \bibnamefont{and}
  \bibinfo{author}{\bibfnamefont{K.}~\bibnamefont{Goeke}},
  \bibinfo{journal}{JHEP} \textbf{\bibinfo{volume}{08}}, \bibinfo{pages}{038}
  (\bibinfo{year}{2008}), \eprint{0805.3165}.

\bibitem[{\citenamefont{Hatta et~al.}(2016)\citenamefont{Hatta, Xiao, and
  Yuan}}]{Hatta:2016dxp}
\bibinfo{author}{\bibfnamefont{Y.}~\bibnamefont{Hatta}},
  \bibinfo{author}{\bibfnamefont{B.-W.} \bibnamefont{Xiao}}, \bibnamefont{and}
  \bibinfo{author}{\bibfnamefont{F.}~\bibnamefont{Yuan}},
  \bibinfo{journal}{Phys. Rev. Lett.} \textbf{\bibinfo{volume}{116}},
  \bibinfo{pages}{202301} (\bibinfo{year}{2016}), \eprint{1601.01585}.

\bibitem[{\citenamefont{Altinoluk et~al.}(2016)\citenamefont{Altinoluk,
  Armesto, Beuf, and Rezaeian}}]{Altinoluk:2015dpi}
\bibinfo{author}{\bibfnamefont{T.}~\bibnamefont{Altinoluk}},
  \bibinfo{author}{\bibfnamefont{N.}~\bibnamefont{Armesto}},
  \bibinfo{author}{\bibfnamefont{G.}~\bibnamefont{Beuf}}, \bibnamefont{and}
  \bibinfo{author}{\bibfnamefont{A.~H.} \bibnamefont{Rezaeian}},
  \bibinfo{journal}{Phys. Lett. B} \textbf{\bibinfo{volume}{758}},
  \bibinfo{pages}{373} (\bibinfo{year}{2016}), \eprint{1511.07452}.

\bibitem[{\citenamefont{Zhou}(2016)}]{Zhou:2016rnt}
\bibinfo{author}{\bibfnamefont{J.}~\bibnamefont{Zhou}}, \bibinfo{journal}{Phys.
  Rev. D} \textbf{\bibinfo{volume}{94}}, \bibinfo{pages}{114017}
  (\bibinfo{year}{2016}), \eprint{1611.02397}.

\bibitem[{\citenamefont{Hagiwara et~al.}(2017)\citenamefont{Hagiwara, Hatta,
  Pasechnik, Tasevsky, and Teryaev}}]{Hagiwara:2017fye}
\bibinfo{author}{\bibfnamefont{Y.}~\bibnamefont{Hagiwara}},
  \bibinfo{author}{\bibfnamefont{Y.}~\bibnamefont{Hatta}},
  \bibinfo{author}{\bibfnamefont{R.}~\bibnamefont{Pasechnik}},
  \bibinfo{author}{\bibfnamefont{M.}~\bibnamefont{Tasevsky}}, \bibnamefont{and}
  \bibinfo{author}{\bibfnamefont{O.}~\bibnamefont{Teryaev}},
  \bibinfo{journal}{Phys. Rev. D} \textbf{\bibinfo{volume}{96}},
  \bibinfo{pages}{034009} (\bibinfo{year}{2017}), \eprint{1706.01765}.

\bibitem[{\citenamefont{Iancu and Rezaeian}(2017)}]{Iancu:2017fzn}
\bibinfo{author}{\bibfnamefont{E.}~\bibnamefont{Iancu}} \bibnamefont{and}
  \bibinfo{author}{\bibfnamefont{A.~H.} \bibnamefont{Rezaeian}},
  \bibinfo{journal}{Phys. Rev. D} \textbf{\bibinfo{volume}{95}},
  \bibinfo{pages}{094003} (\bibinfo{year}{2017}), \eprint{1702.03943}.

\bibitem[{\citenamefont{Boussarie et~al.}(2018)\citenamefont{Boussarie, Hatta,
  Xiao, and Yuan}}]{Boussarie:2018zwg}
\bibinfo{author}{\bibfnamefont{R.}~\bibnamefont{Boussarie}},
  \bibinfo{author}{\bibfnamefont{Y.}~\bibnamefont{Hatta}},
  \bibinfo{author}{\bibfnamefont{B.-W.} \bibnamefont{Xiao}}, \bibnamefont{and}
  \bibinfo{author}{\bibfnamefont{F.}~\bibnamefont{Yuan}},
  \bibinfo{journal}{Phys. Rev. D} \textbf{\bibinfo{volume}{98}},
  \bibinfo{pages}{074015} (\bibinfo{year}{2018}), \eprint{1807.08697}.

\bibitem[{\citenamefont{Hatta et~al.}(2017{\natexlab{b}})\citenamefont{Hatta,
  Xiao, and Yuan}}]{Hatta:2017cte}
\bibinfo{author}{\bibfnamefont{Y.}~\bibnamefont{Hatta}},
  \bibinfo{author}{\bibfnamefont{B.-W.} \bibnamefont{Xiao}}, \bibnamefont{and}
  \bibinfo{author}{\bibfnamefont{F.}~\bibnamefont{Yuan}},
  \bibinfo{journal}{Phys. Rev. D} \textbf{\bibinfo{volume}{95}},
  \bibinfo{pages}{114026} (\bibinfo{year}{2017}{\natexlab{b}}),
  \eprint{1703.02085}.

\bibitem[{\citenamefont{M\"antysaari et~al.}(2019)\citenamefont{M\"antysaari,
  Mueller, and Schenke}}]{Mantysaari:2019csc}
\bibinfo{author}{\bibfnamefont{H.}~\bibnamefont{M\"antysaari}},
  \bibinfo{author}{\bibfnamefont{N.}~\bibnamefont{Mueller}}, \bibnamefont{and}
  \bibinfo{author}{\bibfnamefont{B.}~\bibnamefont{Schenke}},
  \bibinfo{journal}{Phys. Rev. D} \textbf{\bibinfo{volume}{99}},
  \bibinfo{pages}{074004} (\bibinfo{year}{2019}), \eprint{1902.05087}.

\bibitem[{\citenamefont{M\"antysaari et~al.}(2020)\citenamefont{M\"antysaari,
  Mueller, Salazar, and Schenke}}]{Mantysaari:2019hkq}
\bibinfo{author}{\bibfnamefont{H.}~\bibnamefont{M\"antysaari}},
  \bibinfo{author}{\bibfnamefont{N.}~\bibnamefont{Mueller}},
  \bibinfo{author}{\bibfnamefont{F.}~\bibnamefont{Salazar}}, \bibnamefont{and}
  \bibinfo{author}{\bibfnamefont{B.}~\bibnamefont{Schenke}},
  \bibinfo{journal}{Phys. Rev. Lett.} \textbf{\bibinfo{volume}{124}},
  \bibinfo{pages}{112301} (\bibinfo{year}{2020}), \eprint{1912.05586}.

\bibitem[{\citenamefont{M\"antysaari et~al.}(2021)\citenamefont{M\"antysaari,
  Roy, Salazar, and Schenke}}]{Mantysaari:2020lhf}
\bibinfo{author}{\bibfnamefont{H.}~\bibnamefont{M\"antysaari}},
  \bibinfo{author}{\bibfnamefont{K.}~\bibnamefont{Roy}},
  \bibinfo{author}{\bibfnamefont{F.}~\bibnamefont{Salazar}}, \bibnamefont{and}
  \bibinfo{author}{\bibfnamefont{B.}~\bibnamefont{Schenke}},
  \bibinfo{journal}{Phys. Rev. D} \textbf{\bibinfo{volume}{103}},
  \bibinfo{pages}{094026} (\bibinfo{year}{2021}), \eprint{2011.02464}.

\bibitem[{\citenamefont{Dumitru et~al.}(2021)\citenamefont{Dumitru,
  M\"antysaari, Paatelainen, Roy, Salazar, and Schenke}}]{Dumitru:2021mab}
\bibinfo{author}{\bibfnamefont{A.}~\bibnamefont{Dumitru}},
  \bibinfo{author}{\bibfnamefont{H.}~\bibnamefont{M\"antysaari}},
  \bibinfo{author}{\bibfnamefont{R.}~\bibnamefont{Paatelainen}},
  \bibinfo{author}{\bibfnamefont{K.}~\bibnamefont{Roy}},
  \bibinfo{author}{\bibfnamefont{F.}~\bibnamefont{Salazar}}, \bibnamefont{and}
  \bibinfo{author}{\bibfnamefont{B.}~\bibnamefont{Schenke}}, in
  \emph{\bibinfo{booktitle}{{28th International Workshop on Deep Inelastic
  Scattering and Related Subjects}}} (\bibinfo{year}{2021}),
  \eprint{2105.10144}.

\bibitem[{\citenamefont{Bhattacharya et~al.}(2018)\citenamefont{Bhattacharya,
  Metz, Ojha, Tsai, and Zhou}}]{Bhattacharya:2018lgm}
\bibinfo{author}{\bibfnamefont{S.}~\bibnamefont{Bhattacharya}},
  \bibinfo{author}{\bibfnamefont{A.}~\bibnamefont{Metz}},
  \bibinfo{author}{\bibfnamefont{V.~K.} \bibnamefont{Ojha}},
  \bibinfo{author}{\bibfnamefont{J.-Y.} \bibnamefont{Tsai}}, \bibnamefont{and}
  \bibinfo{author}{\bibfnamefont{J.}~\bibnamefont{Zhou}}
  (\bibinfo{year}{2018}), \eprint{1802.10550}.

\bibitem[{\citenamefont{Bhattacharya et~al.}(2017)\citenamefont{Bhattacharya,
  Metz, and Zhou}}]{Bhattacharya:2017bvs}
\bibinfo{author}{\bibfnamefont{S.}~\bibnamefont{Bhattacharya}},
  \bibinfo{author}{\bibfnamefont{A.}~\bibnamefont{Metz}}, \bibnamefont{and}
  \bibinfo{author}{\bibfnamefont{J.}~\bibnamefont{Zhou}},
  \bibinfo{journal}{Phys. Lett. B} \textbf{\bibinfo{volume}{771}},
  \bibinfo{pages}{396} (\bibinfo{year}{2017}), \bibinfo{note}{[Erratum:
  Phys.Lett.B 810, 135866 (2020)]}, \eprint{1702.04387}.

\bibitem[{\citenamefont{Catani et~al.}(2014)\citenamefont{Catani, Grazzini, and
  Torre}}]{Catani:2014qha}
\bibinfo{author}{\bibfnamefont{S.}~\bibnamefont{Catani}},
  \bibinfo{author}{\bibfnamefont{M.}~\bibnamefont{Grazzini}}, \bibnamefont{and}
  \bibinfo{author}{\bibfnamefont{A.}~\bibnamefont{Torre}},
  \bibinfo{journal}{Nucl. Phys. B} \textbf{\bibinfo{volume}{890}},
  \bibinfo{pages}{518} (\bibinfo{year}{2014}), \eprint{1408.4564}.

\bibitem[{\citenamefont{Catani et~al.}(2017)\citenamefont{Catani, Grazzini, and
  Sargsyan}}]{Catani:2017tuc}
\bibinfo{author}{\bibfnamefont{S.}~\bibnamefont{Catani}},
  \bibinfo{author}{\bibfnamefont{M.}~\bibnamefont{Grazzini}}, \bibnamefont{and}
  \bibinfo{author}{\bibfnamefont{H.}~\bibnamefont{Sargsyan}},
  \bibinfo{journal}{JHEP} \textbf{\bibinfo{volume}{06}}, \bibinfo{pages}{017}
  (\bibinfo{year}{2017}), \eprint{1703.08468}.

\bibitem[{\citenamefont{Hatta et~al.}(2021{\natexlab{a}})\citenamefont{Hatta,
  Xiao, Yuan, and Zhou}}]{Hatta:2020bgy}
\bibinfo{author}{\bibfnamefont{Y.}~\bibnamefont{Hatta}},
  \bibinfo{author}{\bibfnamefont{B.-W.} \bibnamefont{Xiao}},
  \bibinfo{author}{\bibfnamefont{F.}~\bibnamefont{Yuan}}, \bibnamefont{and}
  \bibinfo{author}{\bibfnamefont{J.}~\bibnamefont{Zhou}},
  \bibinfo{journal}{Phys. Rev. Lett.} \textbf{\bibinfo{volume}{126}},
  \bibinfo{pages}{142001} (\bibinfo{year}{2021}{\natexlab{a}}),
  \eprint{2010.10774}.

\bibitem[{\citenamefont{Hatta et~al.}(2021{\natexlab{b}})\citenamefont{Hatta,
  Xiao, Yuan, and Zhou}}]{hatta2021azimuthal}
\bibinfo{author}{\bibfnamefont{Y.}~\bibnamefont{Hatta}},
  \bibinfo{author}{\bibfnamefont{B.-W.} \bibnamefont{Xiao}},
  \bibinfo{author}{\bibfnamefont{F.}~\bibnamefont{Yuan}}, \bibnamefont{and}
  \bibinfo{author}{\bibfnamefont{J.}~\bibnamefont{Zhou}}
  (\bibinfo{year}{2021}{\natexlab{b}}), \eprint{2106.05307}.

\bibitem[{\citenamefont{CMS}(2020)}]{CMS:2020ekd}
\bibinfo{author}{\bibnamefont{CMS}} (\bibinfo{year}{2020}),
  \eprint{CMS-PAS-HIN-18-011}.

\bibitem[{\citenamefont{Baur et~al.}(1998)\citenamefont{Baur, Hencken, and
  Trautmann}}]{Baur:1998ay}
\bibinfo{author}{\bibfnamefont{G.}~\bibnamefont{Baur}},
  \bibinfo{author}{\bibfnamefont{K.}~\bibnamefont{Hencken}}, \bibnamefont{and}
  \bibinfo{author}{\bibfnamefont{D.}~\bibnamefont{Trautmann}},
  \bibinfo{journal}{J. Phys. G} \textbf{\bibinfo{volume}{24}},
  \bibinfo{pages}{1657} (\bibinfo{year}{1998}), \eprint{hep-ph/9804348}.

\bibitem[{\citenamefont{Klein and Steinberg}(2020)}]{Klein:2020fmr}
\bibinfo{author}{\bibfnamefont{S.}~\bibnamefont{Klein}} \bibnamefont{and}
  \bibinfo{author}{\bibfnamefont{P.}~\bibnamefont{Steinberg}},
  \bibinfo{journal}{Ann. Rev. Nucl. Part. Sci.} \textbf{\bibinfo{volume}{70}},
  \bibinfo{pages}{323} (\bibinfo{year}{2020}), \eprint{2005.01872}.

\bibitem[{\citenamefont{Klein et~al.}(2020{\natexlab{a}})}]{Adam:2020mxg}
\bibinfo{author}{\bibfnamefont{S.}~\bibnamefont{Klein}} \bibnamefont{et~al.},
  in \emph{\bibinfo{booktitle}{{2022 Snowmass Summer Study}}}
  (\bibinfo{year}{2020}{\natexlab{a}}), \eprint{2009.03838}.

\bibitem[{\citenamefont{Hatta et~al.}(2019)\citenamefont{Hatta, Rajan, and
  Yang}}]{Hatta:2019lxo}
\bibinfo{author}{\bibfnamefont{Y.}~\bibnamefont{Hatta}},
  \bibinfo{author}{\bibfnamefont{A.}~\bibnamefont{Rajan}}, \bibnamefont{and}
  \bibinfo{author}{\bibfnamefont{D.-L.} \bibnamefont{Yang}},
  \bibinfo{journal}{Phys. Rev. D} \textbf{\bibinfo{volume}{100}},
  \bibinfo{pages}{014032} (\bibinfo{year}{2019}), \eprint{1906.00894}.

\bibitem[{\citenamefont{Hatta et~al.}(2020)\citenamefont{Hatta, Strikman, Xu,
  and Yuan}}]{Hatta:2019ocp}
\bibinfo{author}{\bibfnamefont{Y.}~\bibnamefont{Hatta}},
  \bibinfo{author}{\bibfnamefont{M.}~\bibnamefont{Strikman}},
  \bibinfo{author}{\bibfnamefont{J.}~\bibnamefont{Xu}}, \bibnamefont{and}
  \bibinfo{author}{\bibfnamefont{F.}~\bibnamefont{Yuan}},
  \bibinfo{journal}{Phys. Lett. B} \textbf{\bibinfo{volume}{803}},
  \bibinfo{pages}{135321} (\bibinfo{year}{2020}), \eprint{1911.11706}.

\bibitem[{\citenamefont{Steinberg}(2021)}]{Steinberg:2021lfm}
\bibinfo{author}{\bibfnamefont{P.~A.} \bibnamefont{Steinberg}}
  (\bibinfo{collaboration}{ALICE, ATLAS, CMS, LHCb, STAR}),
  \bibinfo{journal}{Nucl. Phys. A} \textbf{\bibinfo{volume}{1005}},
  \bibinfo{pages}{122007} (\bibinfo{year}{2021}).

\bibitem[{\citenamefont{Hattori et~al.}(2021)\citenamefont{Hattori, Taya, and
  Yoshida}}]{Hattori:2020htm}
\bibinfo{author}{\bibfnamefont{K.}~\bibnamefont{Hattori}},
  \bibinfo{author}{\bibfnamefont{H.}~\bibnamefont{Taya}}, \bibnamefont{and}
  \bibinfo{author}{\bibfnamefont{S.}~\bibnamefont{Yoshida}},
  \bibinfo{journal}{JHEP} \textbf{\bibinfo{volume}{01}}, \bibinfo{pages}{093}
  (\bibinfo{year}{2021}), \eprint{2010.13492}.

\bibitem[{\citenamefont{Copinger and Pu}(2020)}]{Copinger:2020nyx}
\bibinfo{author}{\bibfnamefont{P.}~\bibnamefont{Copinger}} \bibnamefont{and}
  \bibinfo{author}{\bibfnamefont{S.}~\bibnamefont{Pu}}, \bibinfo{journal}{Int.
  J. Mod. Phys. A} \textbf{\bibinfo{volume}{35}}, \bibinfo{pages}{2030015}
  (\bibinfo{year}{2020}), \eprint{2008.03635}.

\bibitem[{\citenamefont{Strikman}(2008)}]{Strikman:2008zz}
\bibinfo{author}{\bibfnamefont{M.}~\bibnamefont{Strikman}},
  \bibinfo{journal}{Nucl. Phys. B Proc. Suppl.}
  \textbf{\bibinfo{volume}{179-180}}, \bibinfo{pages}{111}
  (\bibinfo{year}{2008}).

\bibitem[{\citenamefont{Li et~al.}(2019)\citenamefont{Li, Zhou, and
  Zhou}}]{Li:2019yzy}
\bibinfo{author}{\bibfnamefont{C.}~\bibnamefont{Li}},
  \bibinfo{author}{\bibfnamefont{J.}~\bibnamefont{Zhou}}, \bibnamefont{and}
  \bibinfo{author}{\bibfnamefont{Y.-J.} \bibnamefont{Zhou}},
  \bibinfo{journal}{Phys. Lett. B} \textbf{\bibinfo{volume}{795}},
  \bibinfo{pages}{576} (\bibinfo{year}{2019}), \eprint{1903.10084}.

\bibitem[{\citenamefont{Li et~al.}(2020)\citenamefont{Li, Zhou, and
  Zhou}}]{Li:2019sin}
\bibinfo{author}{\bibfnamefont{C.}~\bibnamefont{Li}},
  \bibinfo{author}{\bibfnamefont{J.}~\bibnamefont{Zhou}}, \bibnamefont{and}
  \bibinfo{author}{\bibfnamefont{Y.-J.} \bibnamefont{Zhou}},
  \bibinfo{journal}{Phys. Rev. D} \textbf{\bibinfo{volume}{101}},
  \bibinfo{pages}{034015} (\bibinfo{year}{2020}), \eprint{1911.00237}.

\bibitem[{\citenamefont{Xiao et~al.}(2020)\citenamefont{Xiao, Yuan, and
  Zhou}}]{Xiao:2020ddm}
\bibinfo{author}{\bibfnamefont{B.-W.} \bibnamefont{Xiao}},
  \bibinfo{author}{\bibfnamefont{F.}~\bibnamefont{Yuan}}, \bibnamefont{and}
  \bibinfo{author}{\bibfnamefont{J.}~\bibnamefont{Zhou}}
  (\bibinfo{year}{2020}), \eprint{2003.06352}.

\bibitem[{\citenamefont{Adam et~al.}(2019)}]{Adam:2019mby}
\bibinfo{author}{\bibfnamefont{J.}~\bibnamefont{Adam}} \bibnamefont{et~al.}
  (\bibinfo{collaboration}{STAR}) (\bibinfo{year}{2019}), \eprint{1910.12400}.

\bibitem[{\citenamefont{Brandenburg et~al.}(2019)}]{daniel:2019}
\bibinfo{author}{\bibfnamefont{J.~D.} \bibnamefont{Brandenburg}}
  \bibnamefont{et~al.} (\bibinfo{collaboration}{STAR}), \bibinfo{journal}{talk
  presented in Quark Matter 2019, Wuhan, China}  (\bibinfo{year}{2019}).

\bibitem[{\citenamefont{Xing et~al.}(2020)\citenamefont{Xing, Zhang, Zhou, and
  Zhou}}]{Xing:2020hwh}
\bibinfo{author}{\bibfnamefont{H.}~\bibnamefont{Xing}},
  \bibinfo{author}{\bibfnamefont{C.}~\bibnamefont{Zhang}},
  \bibinfo{author}{\bibfnamefont{J.}~\bibnamefont{Zhou}}, \bibnamefont{and}
  \bibinfo{author}{\bibfnamefont{Y.-J.} \bibnamefont{Zhou}},
  \bibinfo{journal}{JHEP} \textbf{\bibinfo{volume}{10}}, \bibinfo{pages}{064}
  (\bibinfo{year}{2020}), \eprint{2006.06206}.

\bibitem[{\citenamefont{Zha et~al.}(2021)\citenamefont{Zha, Brandenburg, Ruan,
  Tang, and Xu}}]{Zha:2020cst}
\bibinfo{author}{\bibfnamefont{W.}~\bibnamefont{Zha}},
  \bibinfo{author}{\bibfnamefont{J.~D.} \bibnamefont{Brandenburg}},
  \bibinfo{author}{\bibfnamefont{L.}~\bibnamefont{Ruan}},
  \bibinfo{author}{\bibfnamefont{Z.}~\bibnamefont{Tang}}, \bibnamefont{and}
  \bibinfo{author}{\bibfnamefont{Z.}~\bibnamefont{Xu}}, \bibinfo{journal}{Phys.
  Rev. D} \textbf{\bibinfo{volume}{103}}, \bibinfo{pages}{033007}
  (\bibinfo{year}{2021}), \eprint{2006.12099}.

\bibitem[{\citenamefont{Brandenburg et~al.}(2021)\citenamefont{Brandenburg,
  Zha, and Xu}}]{Brandenburg:2021lnj}
\bibinfo{author}{\bibfnamefont{J.~D.} \bibnamefont{Brandenburg}},
  \bibinfo{author}{\bibfnamefont{W.}~\bibnamefont{Zha}}, \bibnamefont{and}
  \bibinfo{author}{\bibfnamefont{Z.}~\bibnamefont{Xu}} (\bibinfo{year}{2021}),
  \eprint{2103.16623}.

\bibitem[{\citenamefont{Hagiwara et~al.}(2021)\citenamefont{Hagiwara, Zhang,
  Zhou, and Zhou}}]{Hagiwara:2020juc}
\bibinfo{author}{\bibfnamefont{Y.}~\bibnamefont{Hagiwara}},
  \bibinfo{author}{\bibfnamefont{C.}~\bibnamefont{Zhang}},
  \bibinfo{author}{\bibfnamefont{J.}~\bibnamefont{Zhou}}, \bibnamefont{and}
  \bibinfo{author}{\bibfnamefont{Y.-J.} \bibnamefont{Zhou}},
  \bibinfo{journal}{Phys. Rev. D} \textbf{\bibinfo{volume}{103}},
  \bibinfo{pages}{074013} (\bibinfo{year}{2021}), \eprint{2011.13151}.

\bibitem[{\citenamefont{Soding}(1966)}]{Soding:1965nh}
\bibinfo{author}{\bibfnamefont{P.}~\bibnamefont{Soding}},
  \bibinfo{journal}{Phys. Lett.} \textbf{\bibinfo{volume}{19}},
  \bibinfo{pages}{702} (\bibinfo{year}{1966}).

\bibitem[{\citenamefont{Klusek-Gawenda and
  Szczurek}(2013)}]{Klusek-Gawenda:2013rtu}
\bibinfo{author}{\bibfnamefont{M.}~\bibnamefont{Klusek-Gawenda}}
  \bibnamefont{and} \bibinfo{author}{\bibfnamefont{A.}~\bibnamefont{Szczurek}},
  \bibinfo{journal}{Phys. Rev. C} \textbf{\bibinfo{volume}{87}},
  \bibinfo{pages}{054908} (\bibinfo{year}{2013}), \eprint{1302.4204}.

\bibitem[{\citenamefont{Bolz et~al.}(2015)\citenamefont{Bolz, Ewerz, Maniatis,
  Nachtmann, Sauter, and Sch\"oning}}]{Bolz:2014mya}
\bibinfo{author}{\bibfnamefont{A.}~\bibnamefont{Bolz}},
  \bibinfo{author}{\bibfnamefont{C.}~\bibnamefont{Ewerz}},
  \bibinfo{author}{\bibfnamefont{M.}~\bibnamefont{Maniatis}},
  \bibinfo{author}{\bibfnamefont{O.}~\bibnamefont{Nachtmann}},
  \bibinfo{author}{\bibfnamefont{M.}~\bibnamefont{Sauter}}, \bibnamefont{and}
  \bibinfo{author}{\bibfnamefont{A.}~\bibnamefont{Sch\"oning}},
  \bibinfo{journal}{JHEP} \textbf{\bibinfo{volume}{01}}, \bibinfo{pages}{151}
  (\bibinfo{year}{2015}), \eprint{1409.8483}.

\bibitem[{\citenamefont{H\"agler et~al.}(2002)\citenamefont{H\"agler, Pire,
  Szymanowski, and Teryaev}}]{Hagler:2002nh}
\bibinfo{author}{\bibfnamefont{P.}~\bibnamefont{H\"agler}},
  \bibinfo{author}{\bibfnamefont{B.}~\bibnamefont{Pire}},
  \bibinfo{author}{\bibfnamefont{L.}~\bibnamefont{Szymanowski}},
  \bibnamefont{and} \bibinfo{author}{\bibfnamefont{O.~V.}
  \bibnamefont{Teryaev}}, \bibinfo{journal}{Phys. Lett. B}
  \textbf{\bibinfo{volume}{535}}, \bibinfo{pages}{117} (\bibinfo{year}{2002}),
  \bibinfo{note}{[Erratum: Phys.Lett.B 540, 324--325 (2002)]},
  \eprint{hep-ph/0202231}.

\bibitem[{\citenamefont{Hagler et~al.}(2002)\citenamefont{Hagler, Pire,
  Szymanowski, and Teryaev}}]{Hagler:2002nf}
\bibinfo{author}{\bibfnamefont{P.}~\bibnamefont{Hagler}},
  \bibinfo{author}{\bibfnamefont{B.}~\bibnamefont{Pire}},
  \bibinfo{author}{\bibfnamefont{L.}~\bibnamefont{Szymanowski}},
  \bibnamefont{and} \bibinfo{author}{\bibfnamefont{O.~V.}
  \bibnamefont{Teryaev}}, \bibinfo{journal}{Eur. Phys. J. C}
  \textbf{\bibinfo{volume}{26}}, \bibinfo{pages}{261} (\bibinfo{year}{2002}),
  \eprint{hep-ph/0207224}.

\bibitem[{\citenamefont{Adamczyk et~al.}(2017)}]{Adamczyk:2017vfu}
\bibinfo{author}{\bibfnamefont{L.}~\bibnamefont{Adamczyk}} \bibnamefont{et~al.}
  (\bibinfo{collaboration}{STAR}), \bibinfo{journal}{Phys. Rev. C}
  \textbf{\bibinfo{volume}{96}}, \bibinfo{pages}{054904}
  (\bibinfo{year}{2017}), \eprint{1702.07705}.

\bibitem[{\citenamefont{Sirunyan et~al.}(2019)}]{Sirunyan:2019nog}
\bibinfo{author}{\bibfnamefont{A.~M.} \bibnamefont{Sirunyan}}
  \bibnamefont{et~al.} (\bibinfo{collaboration}{CMS}), \bibinfo{journal}{Eur.
  Phys. J. C} \textbf{\bibinfo{volume}{79}}, \bibinfo{pages}{702}
  (\bibinfo{year}{2019}), \eprint{1902.01339}.

\bibitem[{\citenamefont{Acharya et~al.}(2020)}]{Acharya:2020sbc}
\bibinfo{author}{\bibfnamefont{S.}~\bibnamefont{Acharya}} \bibnamefont{et~al.}
  (\bibinfo{collaboration}{ALICE}), \bibinfo{journal}{JHEP}
  \textbf{\bibinfo{volume}{06}}, \bibinfo{pages}{035} (\bibinfo{year}{2020}),
  \eprint{2002.10897}.

\bibitem[{\citenamefont{Ryskin}(1993)}]{Ryskin:1992ui}
\bibinfo{author}{\bibfnamefont{M.}~\bibnamefont{Ryskin}}, \bibinfo{journal}{Z.
  Phys. C} \textbf{\bibinfo{volume}{57}}, \bibinfo{pages}{89}
  (\bibinfo{year}{1993}).

\bibitem[{\citenamefont{Brodsky et~al.}(1994)\citenamefont{Brodsky, Frankfurt,
  Gunion, Mueller, and Strikman}}]{Brodsky:1994kf}
\bibinfo{author}{\bibfnamefont{S.~J.} \bibnamefont{Brodsky}},
  \bibinfo{author}{\bibfnamefont{L.}~\bibnamefont{Frankfurt}},
  \bibinfo{author}{\bibfnamefont{J.}~\bibnamefont{Gunion}},
  \bibinfo{author}{\bibfnamefont{A.~H.} \bibnamefont{Mueller}},
  \bibnamefont{and} \bibinfo{author}{\bibfnamefont{M.}~\bibnamefont{Strikman}},
  \bibinfo{journal}{Phys. Rev. D} \textbf{\bibinfo{volume}{50}},
  \bibinfo{pages}{3134} (\bibinfo{year}{1994}), \eprint{hep-ph/9402283}.

\bibitem[{\citenamefont{Bartels et~al.}(2003)\citenamefont{Bartels,
  Golec-Biernat, and Peters}}]{Bartels:2003yj}
\bibinfo{author}{\bibfnamefont{J.}~\bibnamefont{Bartels}},
  \bibinfo{author}{\bibfnamefont{K.~J.} \bibnamefont{Golec-Biernat}},
  \bibnamefont{and} \bibinfo{author}{\bibfnamefont{K.}~\bibnamefont{Peters}},
  \bibinfo{journal}{Acta Phys. Polon. B} \textbf{\bibinfo{volume}{34}},
  \bibinfo{pages}{3051} (\bibinfo{year}{2003}), \eprint{hep-ph/0301192}.

\bibitem[{\citenamefont{Hagiwara et~al.}(2020)\citenamefont{Hagiwara, Hatta,
  Pasechnik, and Zhou}}]{Hagiwara:2020mqb}
\bibinfo{author}{\bibfnamefont{Y.}~\bibnamefont{Hagiwara}},
  \bibinfo{author}{\bibfnamefont{Y.}~\bibnamefont{Hatta}},
  \bibinfo{author}{\bibfnamefont{R.}~\bibnamefont{Pasechnik}},
  \bibnamefont{and} \bibinfo{author}{\bibfnamefont{J.}~\bibnamefont{Zhou}},
  \bibinfo{journal}{Eur. Phys. J. C} \textbf{\bibinfo{volume}{80}},
  \bibinfo{pages}{427} (\bibinfo{year}{2020}), \eprint{2003.03680}.

\bibitem[{\citenamefont{Klein et~al.}(2019)\citenamefont{Klein, Mueller, Xiao,
  and Yuan}}]{Klein:2018fmp}
\bibinfo{author}{\bibfnamefont{S.}~\bibnamefont{Klein}},
  \bibinfo{author}{\bibfnamefont{A.}~\bibnamefont{Mueller}},
  \bibinfo{author}{\bibfnamefont{B.-W.} \bibnamefont{Xiao}}, \bibnamefont{and}
  \bibinfo{author}{\bibfnamefont{F.}~\bibnamefont{Yuan}},
  \bibinfo{journal}{Phys. Rev. Lett.} \textbf{\bibinfo{volume}{122}},
  \bibinfo{pages}{132301} (\bibinfo{year}{2019}), \eprint{1811.05519}.

\bibitem[{\citenamefont{Diehl et~al.}(1998)\citenamefont{Diehl, Gousset, Pire,
  and Teryaev}}]{Diehl:1998dk}
\bibinfo{author}{\bibfnamefont{M.}~\bibnamefont{Diehl}},
  \bibinfo{author}{\bibfnamefont{T.}~\bibnamefont{Gousset}},
  \bibinfo{author}{\bibfnamefont{B.}~\bibnamefont{Pire}}, \bibnamefont{and}
  \bibinfo{author}{\bibfnamefont{O.}~\bibnamefont{Teryaev}},
  \bibinfo{journal}{Phys. Rev. Lett.} \textbf{\bibinfo{volume}{81}},
  \bibinfo{pages}{1782} (\bibinfo{year}{1998}), \eprint{hep-ph/9805380}.

\bibitem[{\citenamefont{Polyakov}(1999)}]{Polyakov:1998ze}
\bibinfo{author}{\bibfnamefont{M.~V.} \bibnamefont{Polyakov}},
  \bibinfo{journal}{Nucl. Phys. B} \textbf{\bibinfo{volume}{555}},
  \bibinfo{pages}{231} (\bibinfo{year}{1999}), \eprint{hep-ph/9809483}.

\bibitem[{\citenamefont{Diehl et~al.}(2000)\citenamefont{Diehl, Gousset, and
  Pire}}]{Diehl:2000uv}
\bibinfo{author}{\bibfnamefont{M.}~\bibnamefont{Diehl}},
  \bibinfo{author}{\bibfnamefont{T.}~\bibnamefont{Gousset}}, \bibnamefont{and}
  \bibinfo{author}{\bibfnamefont{B.}~\bibnamefont{Pire}},
  \bibinfo{journal}{Phys. Rev. D} \textbf{\bibinfo{volume}{62}},
  \bibinfo{pages}{073014} (\bibinfo{year}{2000}), \eprint{hep-ph/0003233}.

\bibitem[{\citenamefont{Pire and Szymanowski}(2003)}]{Pire:2002ut}
\bibinfo{author}{\bibfnamefont{B.}~\bibnamefont{Pire}} \bibnamefont{and}
  \bibinfo{author}{\bibfnamefont{L.}~\bibnamefont{Szymanowski}},
  \bibinfo{journal}{Phys. Lett. B} \textbf{\bibinfo{volume}{556}},
  \bibinfo{pages}{129} (\bibinfo{year}{2003}), \eprint{hep-ph/0212296}.

\bibitem[{\citenamefont{Diehl}(2003)}]{Diehl:2003ny}
\bibinfo{author}{\bibfnamefont{M.}~\bibnamefont{Diehl}},
  \bibinfo{journal}{Phys. Rept.} \textbf{\bibinfo{volume}{388}},
  \bibinfo{pages}{41} (\bibinfo{year}{2003}), \eprint{hep-ph/0307382}.

\bibitem[{\citenamefont{McLerran and
  Venugopalan}(1994{\natexlab{a}})}]{McLerran:1993ka}
\bibinfo{author}{\bibfnamefont{L.~D.} \bibnamefont{McLerran}} \bibnamefont{and}
  \bibinfo{author}{\bibfnamefont{R.}~\bibnamefont{Venugopalan}},
  \bibinfo{journal}{Phys. Rev. D} \textbf{\bibinfo{volume}{49}},
  \bibinfo{pages}{3352} (\bibinfo{year}{1994}{\natexlab{a}}),
  \eprint{hep-ph/9311205}.

\bibitem[{\citenamefont{McLerran and
  Venugopalan}(1994{\natexlab{b}})}]{McLerran:1993ni}
\bibinfo{author}{\bibfnamefont{L.~D.} \bibnamefont{McLerran}} \bibnamefont{and}
  \bibinfo{author}{\bibfnamefont{R.}~\bibnamefont{Venugopalan}},
  \bibinfo{journal}{Phys. Rev. D} \textbf{\bibinfo{volume}{49}},
  \bibinfo{pages}{2233} (\bibinfo{year}{1994}{\natexlab{b}}),
  \eprint{hep-ph/9309289}.

\bibitem[{\citenamefont{Klein and Nystrand}(2000)}]{Klein:1999gv}
\bibinfo{author}{\bibfnamefont{S.~R.} \bibnamefont{Klein}} \bibnamefont{and}
  \bibinfo{author}{\bibfnamefont{J.}~\bibnamefont{Nystrand}},
  \bibinfo{journal}{Phys. Rev. Lett.} \textbf{\bibinfo{volume}{84}},
  \bibinfo{pages}{2330} (\bibinfo{year}{2000}), \eprint{hep-ph/9909237}.

\bibitem[{\citenamefont{Abelev et~al.}(2009)}]{Abelev:2008ew}
\bibinfo{author}{\bibfnamefont{B.}~\bibnamefont{Abelev}} \bibnamefont{et~al.}
  (\bibinfo{collaboration}{STAR}), \bibinfo{journal}{Phys. Rev. Lett.}
  \textbf{\bibinfo{volume}{102}}, \bibinfo{pages}{112301}
  (\bibinfo{year}{2009}), \eprint{0812.1063}.

\bibitem[{\citenamefont{Zha et~al.}(2019)\citenamefont{Zha, Ruan, Tang, Xu, and
  Yang}}]{Zha:2018jin}
\bibinfo{author}{\bibfnamefont{W.}~\bibnamefont{Zha}},
  \bibinfo{author}{\bibfnamefont{L.}~\bibnamefont{Ruan}},
  \bibinfo{author}{\bibfnamefont{Z.}~\bibnamefont{Tang}},
  \bibinfo{author}{\bibfnamefont{Z.}~\bibnamefont{Xu}}, \bibnamefont{and}
  \bibinfo{author}{\bibfnamefont{S.}~\bibnamefont{Yang}},
  \bibinfo{journal}{Phys. Rev. C} \textbf{\bibinfo{volume}{99}},
  \bibinfo{pages}{061901} (\bibinfo{year}{2019}), \eprint{1810.10694}.

\bibitem[{\citenamefont{Chen et~al.}(2021{\natexlab{a}})\citenamefont{Chen,
  Moult, and Zhu}}]{Chen:2020adz}
\bibinfo{author}{\bibfnamefont{H.}~\bibnamefont{Chen}},
  \bibinfo{author}{\bibfnamefont{I.}~\bibnamefont{Moult}}, \bibnamefont{and}
  \bibinfo{author}{\bibfnamefont{H.~X.} \bibnamefont{Zhu}},
  \bibinfo{journal}{Phys. Rev. Lett.} \textbf{\bibinfo{volume}{126}},
  \bibinfo{pages}{112003} (\bibinfo{year}{2021}{\natexlab{a}}),
  \eprint{2011.02492}.

\bibitem[{\citenamefont{Chen et~al.}(2021{\natexlab{b}})\citenamefont{Chen,
  Moult, and Zhu}}]{Chen:2021gdk}
\bibinfo{author}{\bibfnamefont{H.}~\bibnamefont{Chen}},
  \bibinfo{author}{\bibfnamefont{I.}~\bibnamefont{Moult}}, \bibnamefont{and}
  \bibinfo{author}{\bibfnamefont{H.~X.} \bibnamefont{Zhu}}
  (\bibinfo{year}{2021}{\natexlab{b}}), \eprint{2104.00009}.

\bibitem[{\citenamefont{Vidovic et~al.}(1993)\citenamefont{Vidovic, Greiner,
  Best, and Soff}}]{Vidovic:1992ik}
\bibinfo{author}{\bibfnamefont{M.}~\bibnamefont{Vidovic}},
  \bibinfo{author}{\bibfnamefont{M.}~\bibnamefont{Greiner}},
  \bibinfo{author}{\bibfnamefont{C.}~\bibnamefont{Best}}, \bibnamefont{and}
  \bibinfo{author}{\bibfnamefont{G.}~\bibnamefont{Soff}},
  \bibinfo{journal}{Phys. Rev. C} \textbf{\bibinfo{volume}{47}},
  \bibinfo{pages}{2308} (\bibinfo{year}{1993}).

\bibitem[{\citenamefont{Zha et~al.}(2020)\citenamefont{Zha, Brandenburg, Tang,
  and Xu}}]{Zha:2018tlq}
\bibinfo{author}{\bibfnamefont{W.}~\bibnamefont{Zha}},
  \bibinfo{author}{\bibfnamefont{J.~D.} \bibnamefont{Brandenburg}},
  \bibinfo{author}{\bibfnamefont{Z.}~\bibnamefont{Tang}}, \bibnamefont{and}
  \bibinfo{author}{\bibfnamefont{Z.}~\bibnamefont{Xu}}, \bibinfo{journal}{Phys.
  Lett. B} \textbf{\bibinfo{volume}{800}}, \bibinfo{pages}{135089}
  (\bibinfo{year}{2020}), \eprint{1812.02820}.

\bibitem[{\citenamefont{Klein et~al.}(2020{\natexlab{b}})\citenamefont{Klein,
  Mueller, Xiao, and Yuan}}]{Klein:2020jom}
\bibinfo{author}{\bibfnamefont{S.}~\bibnamefont{Klein}},
  \bibinfo{author}{\bibfnamefont{A.}~\bibnamefont{Mueller}},
  \bibinfo{author}{\bibfnamefont{B.-W.} \bibnamefont{Xiao}}, \bibnamefont{and}
  \bibinfo{author}{\bibfnamefont{F.}~\bibnamefont{Yuan}}
  (\bibinfo{year}{2020}{\natexlab{b}}), \eprint{2003.02947}.

\bibitem[{\citenamefont{K\l{}usek-Gawenda
  et~al.}(2021)\citenamefont{K\l{}usek-Gawenda, Sch\"afer, and
  Szczurek}}]{Klusek-Gawenda:2020eja}
\bibinfo{author}{\bibfnamefont{M.}~\bibnamefont{K\l{}usek-Gawenda}},
  \bibinfo{author}{\bibfnamefont{W.}~\bibnamefont{Sch\"afer}},
  \bibnamefont{and} \bibinfo{author}{\bibfnamefont{A.}~\bibnamefont{Szczurek}},
  \bibinfo{journal}{Phys. Lett. B} \textbf{\bibinfo{volume}{814}},
  \bibinfo{pages}{136114} (\bibinfo{year}{2021}), \eprint{2012.11973}.

\bibitem[{\citenamefont{Wu}(2021)}]{Wu:2021ril}
\bibinfo{author}{\bibfnamefont{B.}~\bibnamefont{Wu}} (\bibinfo{year}{2021}),
  \eprint{2102.12916}.

\bibitem[{\citenamefont{Kowalski and Teaney}(2003)}]{Kowalski:2003hm}
\bibinfo{author}{\bibfnamefont{H.}~\bibnamefont{Kowalski}} \bibnamefont{and}
  \bibinfo{author}{\bibfnamefont{D.}~\bibnamefont{Teaney}},
  \bibinfo{journal}{Phys. Rev. D} \textbf{\bibinfo{volume}{68}},
  \bibinfo{pages}{114005} (\bibinfo{year}{2003}), \eprint{hep-ph/0304189}.

\bibitem[{\citenamefont{Kowalski et~al.}(2006)\citenamefont{Kowalski, Motyka,
  and Watt}}]{Kowalski:2006hc}
\bibinfo{author}{\bibfnamefont{H.}~\bibnamefont{Kowalski}},
  \bibinfo{author}{\bibfnamefont{L.}~\bibnamefont{Motyka}}, \bibnamefont{and}
  \bibinfo{author}{\bibfnamefont{G.}~\bibnamefont{Watt}},
  \bibinfo{journal}{Phys. Rev. D} \textbf{\bibinfo{volume}{74}},
  \bibinfo{pages}{074016} (\bibinfo{year}{2006}), \eprint{hep-ph/0606272}.

\bibitem[{\citenamefont{Rezaeian et~al.}(2013)\citenamefont{Rezaeian, Siddikov,
  Van~de Klundert, and Venugopalan}}]{Rezaeian:2012ji}
\bibinfo{author}{\bibfnamefont{A.~H.} \bibnamefont{Rezaeian}},
  \bibinfo{author}{\bibfnamefont{M.}~\bibnamefont{Siddikov}},
  \bibinfo{author}{\bibfnamefont{M.}~\bibnamefont{Van~de Klundert}},
  \bibnamefont{and}
  \bibinfo{author}{\bibfnamefont{R.}~\bibnamefont{Venugopalan}},
  \bibinfo{journal}{Phys. Rev. D} \textbf{\bibinfo{volume}{87}},
  \bibinfo{pages}{034002} (\bibinfo{year}{2013}), \eprint{1212.2974}.

\bibitem[{\citenamefont{Kowalski
  et~al.}(2008{\natexlab{a}})\citenamefont{Kowalski, Lappi, Marquet, and
  Venugopalan}}]{Kowalski:2008sa}
\bibinfo{author}{\bibfnamefont{H.}~\bibnamefont{Kowalski}},
  \bibinfo{author}{\bibfnamefont{T.}~\bibnamefont{Lappi}},
  \bibinfo{author}{\bibfnamefont{C.}~\bibnamefont{Marquet}}, \bibnamefont{and}
  \bibinfo{author}{\bibfnamefont{R.}~\bibnamefont{Venugopalan}},
  \bibinfo{journal}{Phys. Rev. C} \textbf{\bibinfo{volume}{78}},
  \bibinfo{pages}{045201} (\bibinfo{year}{2008}{\natexlab{a}}),
  \eprint{0805.4071}.

\bibitem[{\citenamefont{Kowalski
  et~al.}(2008{\natexlab{b}})\citenamefont{Kowalski, Lappi, and
  Venugopalan}}]{Kowalski:2007rw}
\bibinfo{author}{\bibfnamefont{H.}~\bibnamefont{Kowalski}},
  \bibinfo{author}{\bibfnamefont{T.}~\bibnamefont{Lappi}}, \bibnamefont{and}
  \bibinfo{author}{\bibfnamefont{R.}~\bibnamefont{Venugopalan}},
  \bibinfo{journal}{Phys. Rev. Lett.} \textbf{\bibinfo{volume}{100}},
  \bibinfo{pages}{022303} (\bibinfo{year}{2008}{\natexlab{b}}),
  \eprint{0705.3047}.

\bibitem[{\citenamefont{Bartels et~al.}(2002)\citenamefont{Bartels,
  Golec-Biernat, and Kowalski}}]{Bartels:2002cj}
\bibinfo{author}{\bibfnamefont{J.}~\bibnamefont{Bartels}},
  \bibinfo{author}{\bibfnamefont{K.~J.} \bibnamefont{Golec-Biernat}},
  \bibnamefont{and} \bibinfo{author}{\bibfnamefont{H.}~\bibnamefont{Kowalski}},
  \bibinfo{journal}{Phys. Rev. D} \textbf{\bibinfo{volume}{66}},
  \bibinfo{pages}{014001} (\bibinfo{year}{2002}), \eprint{hep-ph/0203258}.

\end{thebibliography}

\end {document}